\documentclass[sigconf]{acmart}
\usepackage{booktabs} % For formal tables
\usepackage{multirow}
\usepackage{graphicx}
\usepackage{subfigure}
\usepackage{algorithm}
\usepackage{algorithmic}
\usepackage{color}
\usepackage{amsmath}
\usepackage{textcomp}
\usepackage{balance}
\usepackage{bm}
\usepackage[T1]{fontenc}
\AtBeginDocument{%
  \providecommand\BibTeX{{%
    \normalfont B\kern-0.5em{\scshape i\kern-0.25em b}\kern-0.8em\TeX}}}

\copyrightyear{2020}
\acmYear{2020}
\setcopyright{iw3c2w3}
\acmConference[WWW '20]{Proceedings of The Web Conference 2020}{April 20--24, 2020}{Taipei, Taiwan}
\acmBooktitle{Proceedings of The Web Conference 2020 (WWW '20), April 20--24, 2020, Taipei, Taiwan}
\acmPrice{}
\acmDOI{10.1145/3366423.3380050}
\acmISBN{978-1-4503-7023-3/20/04}

\author{Suyu Ge}
\affiliation{%
  \institution{Tsinghua National Laboratory for Information Science and Technology}
  \city{Tsinghua University}
}
\email{gesy17@mails.tsinghua.edu.cn}

\author{Chuhan Wu}
\affiliation{%
  \institution{Tsinghua National Laboratory for Information Science and Technology}
  \city{Tsinghua University}
}
\email{wuchuhan15@gmail.com}

\author{Fangzhao Wu}
\affiliation{%
  \institution{Microsoft Research Asia}
  \city{Beijing}
  \state{China}
}
\email{wufangzhao@gmail.com}

\author{Tao Qi}
\affiliation{%
  \institution{Tsinghua National Laboratory for Information Science and Technology}
  \city{Tsinghua University}
}
\email{qit16@mails.tsinghua.edu.cn}

\author{Yongfeng Huang}
\affiliation{%
  \institution{Tsinghua National Laboratory for Information Science and Technology}
  \city{Tsinghua University}
}
\email{yfhuang@tsinghua.edu.cn}

\renewcommand{\shortauthors}{}

\begin{document}
\title{Graph Enhanced Representation Learning \\for News Recommendation}

\begin{abstract}
With the explosion of online news, personalized news recommendation becomes increasingly important for online news platforms to help their users find interesting information.
Existing news recommendation methods achieve personalization by building accurate news representations from news content and user representations from their direct interactions with news (e.g., click), while ignoring the high-order relatedness between users and news.
Here we propose a news recommendation method which can enhance the representation learning of users and news by modeling their relatedness in a graph setting.
In our method, users and news are both viewed as nodes in a bipartite graph constructed from historical user click behaviors.
For news representations, a transformer architecture is first exploited to build news semantic representations.
Then we combine it with the information from neighbor news in the graph via a graph attention network.
For user representations, we not only represent users from their historically clicked news, but also attentively incorporate the representations of their neighbor users in the graph.
%Experiments were conducted on a large-scale real-world dataset.
%The improved performances validate the effectiveness of our proposed method
Improved performances on a large-scale real-world dataset validate the effectiveness of our proposed method.
\end{abstract}

\begin{CCSXML}
<ccs2012>
<concept>
<concept_id>10002951.10003317.10003347.10003350</concept_id>
<concept_desc>Information systems~Recommender systems</concept_desc>
<concept_significance>500</concept_significance>
</concept>
<concept>
<concept_id>10010147.10010178.10010179</concept_id>
<concept_desc>Computing methodologies~Natural language processing</concept_desc>
<concept_significance>500</concept_significance>
</concept>
</ccs2012>
\end{CCSXML}
\ccsdesc[500]{Information systems~Recommender systems}
\ccsdesc[500]{Computing methodologies~Natural language processing}

\keywords{News Recommendation, Transformer, Graph Attention Network}

\maketitle
\section{Introduction}
Both the overwhelming number of newly-sprung news and huge volumes of online news consumption pose challenges to online news aggregating platforms.
Thus, how to target different users' news reading interests and avoid showcasing excessive irrelevant news becomes an important problem for these platforms~\cite{phelan2011terms,liu2010personalized}.
A possible solution is personalized news recommendation, which depicts user interests from previous user-news interactions~\cite{li2011scene,bansal2015content}.
However, unlike general personalized recommendation methods, news recommendation is unique from certain aspects.
The fast iteration speed of online news makes traditional ID-based recommendation methods such as collaborative filtering (CF) suffer from data sparsity problem~\cite{guo2014merging}.
Meanwhile, rich semantic information in news texts distinguishes itself from recommendation in other domains (e.g., music, fashion and food).
Therefore, a precise understanding of textual content is also vital for news recommendation.
\begin{figure}[t]
    \centering
    \resizebox{0.4\textwidth}{!}{\includegraphics{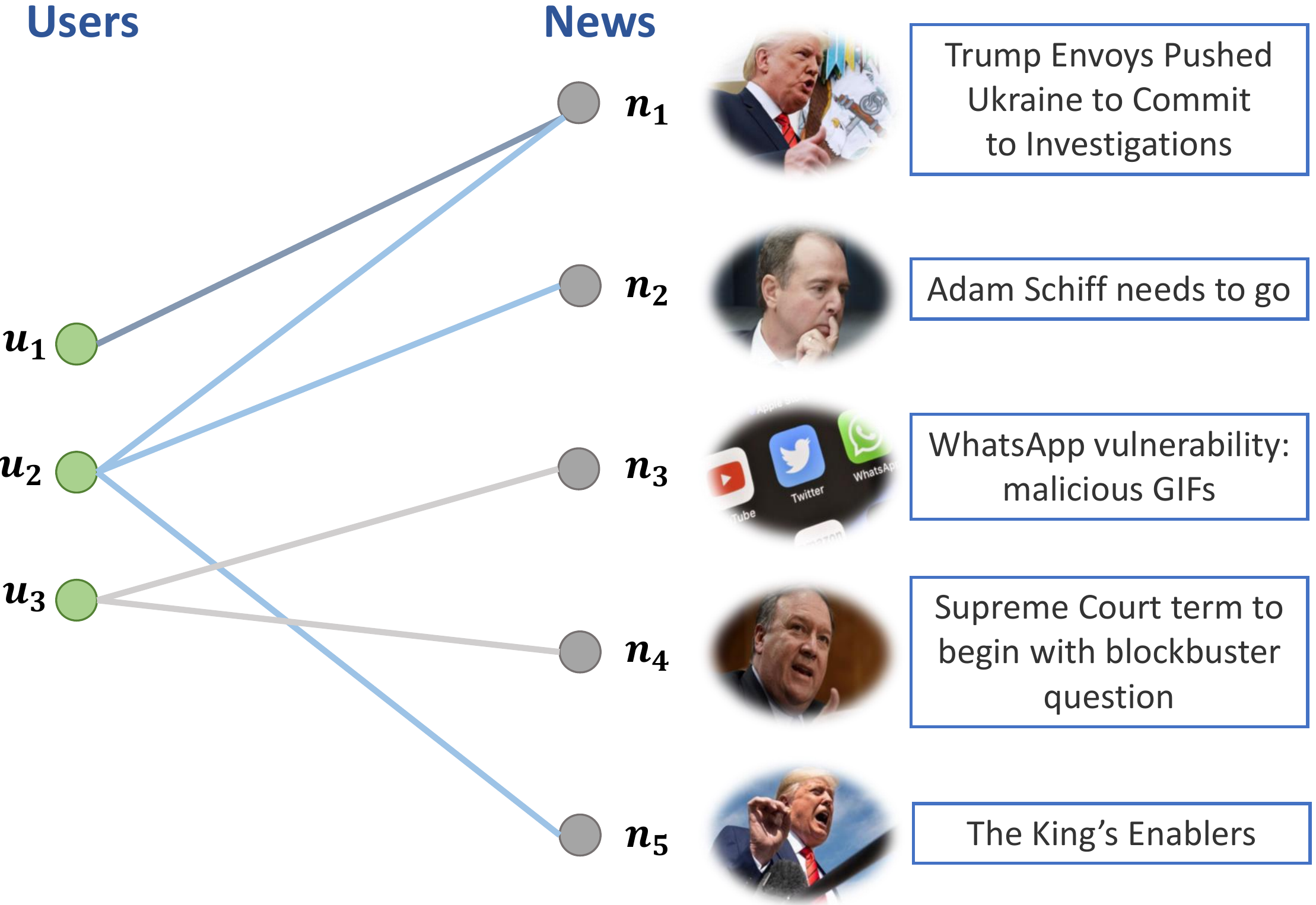}}
    \caption{A user-news bipartite graph.}
    \label{fig:example}
\end{figure}

Existing news recommendation methods achieve personalized news ranking by building accurate news and user representations.
They usually build news representations from news content~\cite{bansal2015content,lian2018towards,zhu2019dan,wwu2019neural}.
Based on that, user representations are constructed from their click behaviors, e.g., the aggregation of their clicked news representations.
For example, Wang et al. proposed DKN~\cite{wang2018dkn}, which formed news representations from their titles via convolutional neural network (CNN).
Then they utilized an attention network to select important clicked news for user representations.
Wu et al.~\cite{wu2019npa} further enhanced personalized news representations by incorporating user IDs as attention queries to select important words in news titles.
When forming user representations, the same attention query was used to select important clicked news.
Compared with traditional collaborative filtering methods~\cite{konstan1997grouplens,ren2017social,ling2014ratings}, which suffer heavy cold-start problems~\cite{lika2014facing}, these methods gained a competitive edge by learning semantic news representations directly from news context.
However, most of them build news representations only from news content and build user representations only from users' historically clicked news.
When the news content such as titles are short and vague, and the historical behaviors of user are sparse, it is difficult for them to learn accurate news and user representations.

Our work is motivated by several observations. 
First, from user-news interactions, a bipartite graph can be established.
Within this graph, both users and news can be viewed as nodes and interactions between them can be viewed as edges.
Among them, some news are viewed by the same user, thus are defined as neighbor news.
Similarly, specific users may share common clicked news, and are denoted as neighbor users.
For example, in Figure~\ref{fig:example}, news $n_1$ and $n_5$ are neighbors because they are both clicked by user $u_2$.
Meanwhile, $u_1$ and $u_2$ are neighbor users.
Second, news representation may be enhanced by considering neighbor news in the graph.
For example, neighbor news $n_1$ and $n_5$ both relates to politics.
However, the expression ``The King'' in $n_5$ is vague without any external information.
By linking it to news $n_1$, which is more detailed and explicit, we may infer that $n_5$ talks about president Trump.
Thus, when forming news representation for $n_5$, $n_1$ may be modeled simultaneously as a form of complementary information.
Third, neighbor users in the graph may share some similar news preferences. 
Incorporating such similarities may further enrich target user representations.
As illustrated, $u_1$ and $u_2$ share common clicked news $n_1$, indicating that they may be both interested in political news.
Nevertheless, it is challenging to form accurate user representation for $u_1$ since the click history of $u_1$ is very sparse.
Thus, explicitly introducing information from $u_2$ may enrich the representation of $u_1$ and lead to better recommendation performances.

In this paper, we propose to incorporate the graph relatedness of users and news to enhance their representation learning for news recommendation.
First, we utilize the transformer~\cite{vaswani2017attention} to build news semantic representations from textual content.
In this way, the multi-head self-attention network encodes word dependency in titles at both short and long distance.
We also add topic embeddings of news since they may contain important information.
Then we further enhance news representations by aggregating neighbor news via a graph attention network.
To enrich neighbour news representations, we utilize both their semantic representations and ID embeddings.
For user representations, besides attentively building user representations from their ID embeddings and historically clicked news, our approach also leverages graph information.
We use the attention mechanism to aggregate the ID embeddings of neighbor users.
Finally, recommendation is made by taking the dot product between user and news representations.
We conduct extensive experiments on a large real-world dataset.
The improved performances over a set of well-known baselines validate the effectiveness of our approach.
\section{Related Work}

Neural news recommendation receives attention from both data mining and natural language processing fields~\cite{zheng2018drn,wang2017dynamic,hamilton2017representation}.
Many previous works handle this problem by learning news and user representations from textual content~\cite{wu2019npa,an-etal-2019-neural,zhu2019dan,wuu2019neural}.
From such viewpoint, user representations are built upon clicked news representations using certain summation techniques (e.g., attentive aggregation or sequential encoding).
For instance, Okura~\cite{okura2017embedding} incorporated denoising autoencoder to form news representations.
Then they explored various types of recurrent networks to encode users.
An et. al~\cite{an-etal-2019-neural} attentively encoded news by combining title and topic information. 
They learned news representations via CNN and formed user representations from their clicked news via a gated recurrent unit (GRU) network.
Zhu et. al.~\cite{zhu2019dan} exploited long short-term memory network (LSTM) to encode clicked news, then applied a single-directional attention network to select important click history for user representations.
Though effective in extracting information from textual content, the works presented above neglect relatedness between neighbor users (or items) in the interaction graph.
Different from their methods, our approach exploits both context meaning and neighbor relatedness in graph.

Recently, graph neural networks (GNN) have received wide attention, and a surge of attempts have been made to develop GNN architectures for recommender systems~\cite{ying2018graph,wu2019session,hamilton2017representation}.
These models leverage both node attributes and graph structure by representing users and items using a combination of neighbor node embeddings~\cite{song2019session}.
For instance, Wang et. al.~\cite{KGAT19} combined knowledge graph (KG) with collaborative signals via a graph attention network, thus enhancing user and item representations with entity information in KG.
Ying et. al.~\cite{ying2018graph} introduced graph convolution to web-scale recommendation.
Node representations of users and items were formed using visual and annotation features.
% However, in most works, the recommendation task is optimized by propagating through the entire graph in a Laplacian matrix form, which may bring considerable computational cost when facing massive news and user nodes~\cite{hamilton2017representation}.
% Moreover, in such a scenario, representations are initially formed via node embedding, then optimized by receiving propagation signals from the graph~\cite{NGCF19}.
% Although their node representations can be enhanced by adding item relation~\cite{RCF}, visual features~\cite{ying2018graph} or knowledge graphs~\cite{Wang2019KGN}, it is still difficult to integrate sophisticated context understanding techniques to the graph learning stage.
% Under this circumstance, rich semantic meanings in the textual content may not be fully exploited. 
% Different from their works, our approach does not operate on the whole adjacency matrix.
% Instead, we propose to optimize only on node pairs (e.g., a user and its neighbor users), thus reducing the excessive computational cost.
% Meanwhile, we leverage the transformer architecture to encode semantic node representations.
% Hence, in our approach, node representations are based on precise context understanding.
In most works, representations are initially formed via node embedding, then optimized by receiving propagation signals from the graph~\cite{NGCF19,wu2019session}.
Although node embeddings are enhanced by adding item relation~\cite{RCF}, visual features~\cite{ying2018graph} or knowledge graphs~\cite{Wang2019KGN}, rich semantic information in the textual content may not be fully exploited.
Different form their work, our approach learns the node embeddings of news directly from its textual content.
We utilize the transformer architecture to model context dependency in news titles.
Thus, our approach improves the node embedding by forming context-aware news representation.
% Meanwhile, in graph-based recommendation approaches, the recommendation task is usually optimized by propagating through the entire graph in a Laplacian matrix form, which may bring considerable computational cost when facing massive news and user nodes~\cite{hamilton2017representation}.
% Conversely, our approach does not propagate on the whole matrix.
% When making recommendation, only a fixed number of neighbor news and neighbor user representations are aggregated, which saves the computational cost by updating the local graph.
\section{Our Approach}

\begin{figure*}[t]
    \centering
    \subfigure[Overview of the model.]{\label{mainModel} \resizebox{0.5\textwidth}{!}{\includegraphics{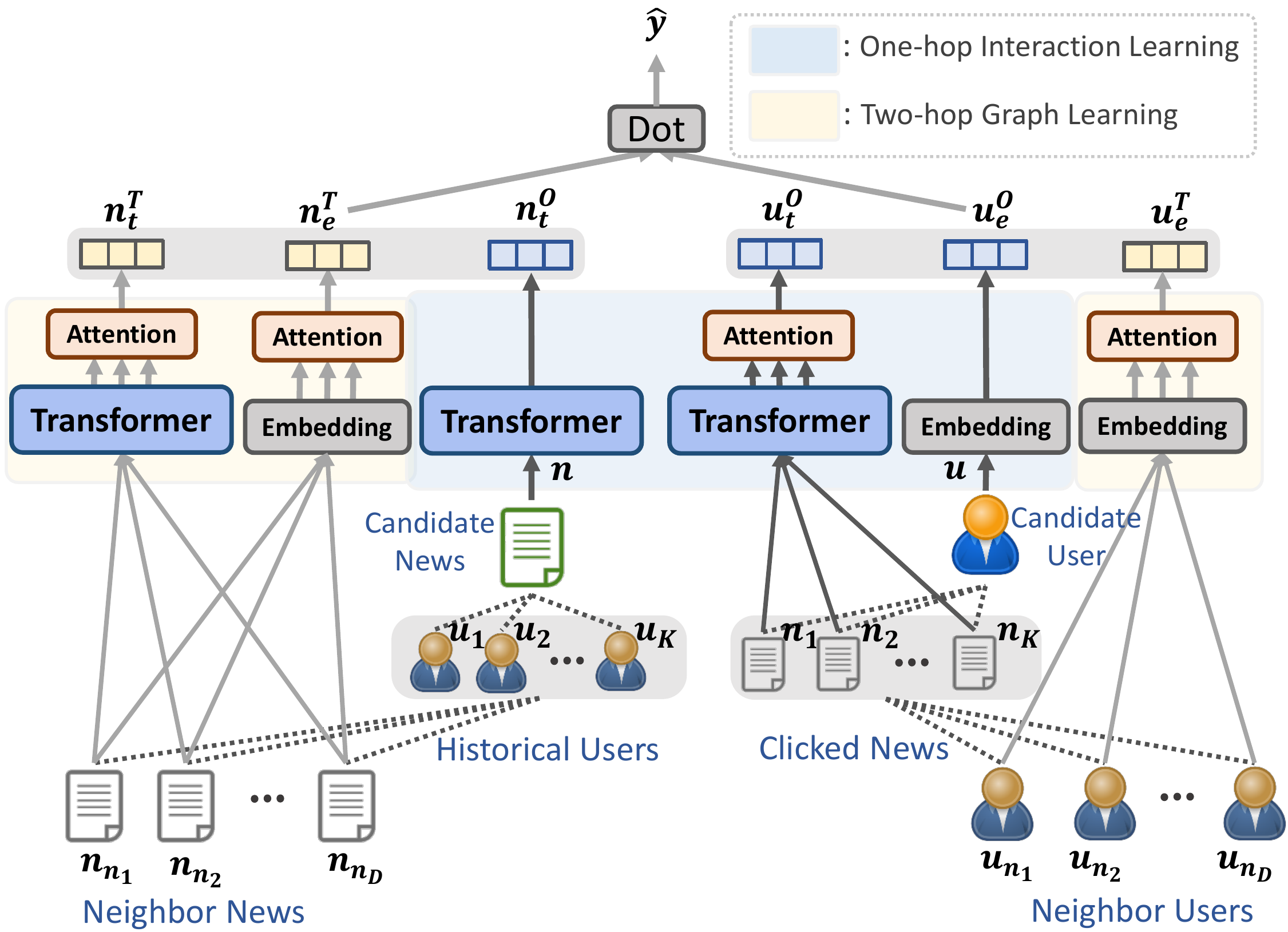}} }
    \qquad \qquad \qquad
    \subfigure[Transformer submodule.]{\label{trans} \resizebox{0.2\textwidth}{!}{\includegraphics{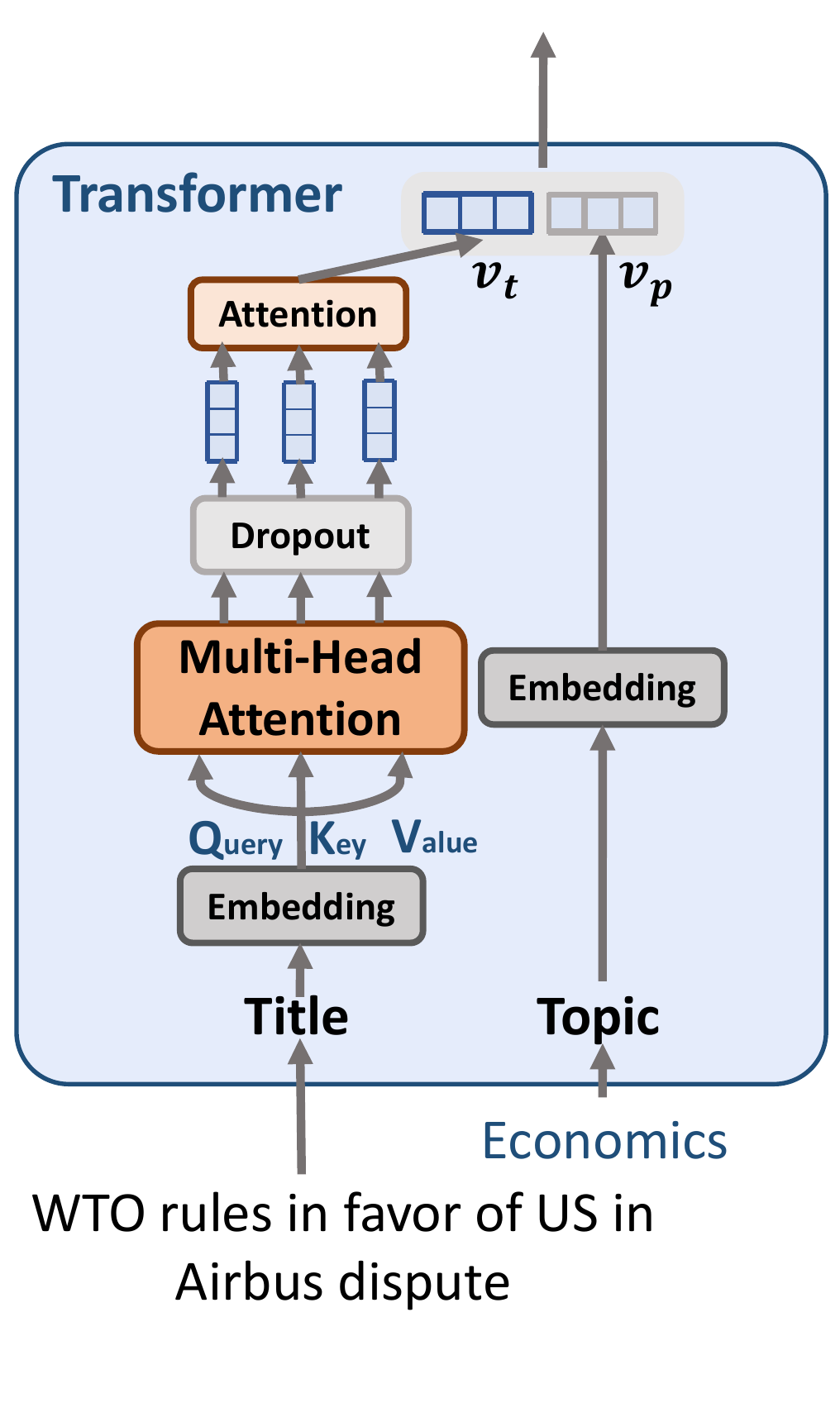}} }
    \caption{An illustration of our proposed GERL approach. 
    Dashed lines represent graph connectivity established from click behaviors, and solid lines represent the information flow among different modules.}
    \label{fig:model}
\end{figure*}

In this section, we will introduce our \textit{\textbf{G}raph \textbf{E}nhanced \textbf{R}epresentation \textbf{L}earning} (\textbf{GERL}) approach illustrated in Figure~\ref{fig:model}, which consists of a \textit{one-hop interaction learning} module and a \textit{two-hop graph learning} module.
The \textit{one-hop interaction learning} module represents target user from historically clicked news and represents candidate news based on its textual content.
% To formulate textual representations, we employ a transformer architecture, which mainly encodes titles via multi-head self-attention.
The \textit{two-hop graph learning} module learns neighbor embeddings of news and users using a graph attention network.
% To enhance neighbor representations from different aspects, we encode neighbors using both textual representations and ID embeddings.

\subsection{Transformer for Context Understanding}
Motivated by Vaswani et al.~\cite{vaswani2017attention}, we utilize the transformer to form accurate context representations from news titles and topics.
News titles are usually clear and concise.
Hence, to avoid the degradation of performance caused by excessive parameters, we simplify the transformer to single layer of multi-head attention.\footnote{We also tried the original transformer architecture but the performance is sub-optimal.}

We then introduce the modified transformer from bottom to top.
The bottom layer is the word embedding, which converts words in a news title into a sequence of low-dimensional embedding vectors.
Denote a news title with $M$ words as $[w_1, w_2, ..., w_M]$, through this layer it is converted into the embedded vector sequence $[\mathbf{e}_1, \mathbf{e}_2, ..., \mathbf{e}_M]$.

The following layer is a word-level multi-head self-attention network.
Interactions between words are important for learning news representations.
For instance, in the title ``Sparks gives Penny Toler a fire from the organization'', the interaction between ``Sparks'' and ``organization'' helps understand the title.
Moreover, a word may relate to more than one words in the title.
For example, the word ``Sparks'' has interactions with both words ``fire'' and ``organization''.
Thus, we employ the multi-head self-attention to form contextual word representations.
The representation of the $i^{th}$ word learned by the $k^{th}$ attention head is computed as:
\begin{equation}
    \begin{split}
        \alpha_{i,j}^k=\frac{\exp(\mathbf{e}_i^T \mathbf{W}^k_s \mathbf{e}_j)}{\sum_{m=1}^M \exp(\mathbf{e}_i^T \mathbf{W}_s^k \mathbf{e}_m)},\\
        \mathbf{h}_{i}^k=\mathbf{W}_v^k(\Sigma_{j=1}^M \alpha_{i,j}^k \mathbf{e}_j),
    \end{split}
\end{equation}
% \begin{equation}
%     \mathbf{h}_{i,k}=\mathbf{W}_v^k(\Sigma_{j=1}^M softmax(\mathbf{e}_i^T \mathbf{W}^k_s \mathbf{e}_j) \mathbf{e}_j),
% \end{equation}
where $\mathbf{W}^k_s$ and $\mathbf{W}_v^k$ are the projection matrices in the $k^{th}$ self-attention head, and $\alpha_{i,j}^k$ indicates the relative importance of the relatedness between the $i^{th}$ and $j^{th}$ words.
The multi-head representation $\mathbf{h}^i$ of the $i^{th}$ word is the concatenation of the representations produced by $N$ separate self-attention heads, i.e., $\mathbf{h}_i=[\mathbf{h}_{i}^1; \mathbf{h}_{i}^2; ...; \mathbf{h}_{i}^N].$
To mitigate overfitting, we add dropout~\cite{srivastava2014dropout} after the self-attention.

Next, we utilize an additive word attention network to model relative importance of different words and aggregate them into title representations.
For instance, the word ``fire'' is more important than other words in the above example.
The attention weight $\beta_i^w$ of the $i_{th}$ word is computed as:
\begin{equation}
    \beta_i^w=\frac{\exp(\mathbf{q}_w^T \tanh(\mathbf{U}_w \times \mathbf{h}_i+\mathbf{u}_w))}{\sum_{j=1}^M\exp(\mathbf{q}_w^T \tanh(\mathbf{U}_w \times \mathbf{h}_j+\mathbf{u}_w))},
\end{equation}
where $\mathbf{q}_w$, $\mathbf{U}_w$ and $\mathbf{u}_w$ are trainable parameters in the word attention network.
The news title representation $\mathbf{v}_t$ is then calculated as: $\mathbf{v}_t= \Sigma_{i=1}^M \beta_i^w\mathbf{h}_i$.

Since topics of user clicked news may also reveal their preferences, we model news topics via an embedding matrix.
Denote the output of this embedding matrix as $\mathbf{v}_p$, then the final representation of the news is the concatenation of the title vector and the topic vector, i.e., $\mathbf{v}=[\mathbf{v}_t; \mathbf{v}_p]$.

\subsection{One-hop Interaction Learning}
The \textit{one-hop interaction learning} module learns candidate news and click behaviors of target users.
More specifically, it can be decomposed into three parts:
(1) Candidate news semantic representations;
(2) Target user semantic representations;
(3) Target user ID representations.

\textbf{Candidate News Semantic Representations.}
Since understanding the content of candidate news is crucial for recommendation, we propose to utilize the transformer to form accurate representation of it.
Given the candidate news $n$, the one-hop (denoted as superscript $^O$) output of the transformer module (denoted as subscript $_t$) is $\mathbf{n}_{t}^{O}$.

\textbf{Target User Semantic Representations.}
The news reading preference of a user can be clearly revealed by their clicked news.
Thus, we propose to model user representations from the content of their clicked news.
Besides, different news may have varied importance for modeling user interests.
For example, the news ``crazy storms hit Los Angeles'' is less important than the news ``6 most popular music dramas'' in modeling user interests.
Thus, we apply an additive attention mechanism to aggregate clicked news vectors for user representations.
Given a target user $u$ and a total number of $K$ clicked news $[n_{1}, n_{2}, ..., n_{K}]$, we first get their transformer encoded outputs $[\mathbf{v}_1, \mathbf{v}_2, ..., \mathbf{v}_K]$.
Then the attention weight $\beta_i^n$ of the $i^{th}$ clicked news is calculated as:
\begin{equation}
     \beta_i^n=\frac{\exp(\mathbf{q}_n^T \tanh(\mathbf{U}_n \times \mathbf{v}_i+\mathbf{u}_n))}{\Sigma_{q=1}^K\exp(\mathbf{q}_n^T \tanh(\mathbf{U}_n \times \mathbf{v}_q+\mathbf{u}_n))},
\end{equation}
where $\mathbf{q}_n$, $\mathbf{U}_n$ and $\mathbf{u}_n$ are the trainable parameters of the news attention network.
The one-hop user semantic representation $\mathbf{u}_{t}^{O}$ is then calculated as: $\mathbf{u}_{t}^{O}= \Sigma_{i=1}^K \beta_i^n \mathbf{v}_i$.

\textbf{Target User ID Representations.}
Since user IDs represent each user uniquely, we incorporate them as latent representations of user interests~\cite{lv2011learning,marlin2004multiple}.
We use a trainable ID embedding matrix $\mathcal{M}_u\in \mathcal{R}^{N_u\times Q}$ to represent each user ID as a low-dimensional vector, where $N_u$ is the number of users and $Q$ is the dimension of the ID embedding. For the user $u$, the one-hop ID embedding vector is denoted as $\mathbf{u}_{e}^{O}$.

\subsection{Two-hop Graph Learning}
\label{sec:graph}
% the motivation towards our graph method should be strengthened.
The \textit{two-hop graph learning} module mines the relatedness between neighbor users and news from the interaction graph.
% Different from typical graph neural network~\cite{velivckovic2017graph,kipf2016semi}, here the inputs are only neighbor node pairs and their associated news content.
% Therefore, it is non-trivial to explicitly encode the graph information.
Additionally, for a given target user, neighbor users usually have different levels of similarity with her/his.
The same situation exists between neighbor news.
To utilize this kind of similarity, we aggregate neighbor news and user information with a graph attention network~\cite{song2019session}.
The utilized graph information here is heterogeneous, including both semantic representations and ID embeddings.
In this \textit{two-hop graph learning} module, there are also three parts:
(1) Neighbor user ID representations;
(2) Neighbor news ID representations;
(3) Neighbor news semantic representations.

\textbf{Neighbor User ID Representations.}
Since adding neighbor user information may complement target user representations, we aggregate the ID embeddings of neighbor users via an additive attention network.
Given a user $u$ and a list of $D$ neighbor users $[u_{n_1}, u_{n_2}, ..., u_{n_D}]$, we first get their ID embeddings via the same user ID embedding matrix 
$\mathcal{M}_u$, which are denoted as $[\mathbf{m}_{u_1}, \mathbf{m}_{u_2}, ..., \mathbf{m}_{u_D}]$.
Then the attention weight $\beta_i^u$ of the $i^{th}$ neighbor user is calculated as:
\begin{equation}
     \beta_i^u=\frac{\exp(\mathbf{q}_u^T \tanh(\mathbf{U}_u \times \mathbf{m}_{u_i}+\mathbf{u}_u))}{\Sigma_{q=1}^D\exp(\mathbf{q}_u^T \tanh(\mathbf{U}_u \times \mathbf{m}_{u_q}+\mathbf{u}_u))},
\label{equ:att}
\end{equation}
where $\mathbf{q}_u$, $\mathbf{U}_u$ and $\mathbf{u}_u$ are trainable parameters in the neighbor user attention network.
The two-hop neighbor user ID representation $\mathbf{u}_{e}^{T}$ is then calculated as: $\mathbf{u}_{e}^{T}= \Sigma_{i=1}^D \beta_i^u \mathbf{m}_{u_i}$.

\textbf{Neighbor News ID Representations.}
News clicked by the same user reveal certain preference of the user, thus may share some common characteristics.
To model this kind of similarity, we utilize an attention network to learn neighbor news ID representations.
For news $n$ with a list of $D$ neighbor news $[n_{n_1}, n_{n_2}, ..., n_{n_D}]$, we first transform neighbors with the news ID embedding matrix $\mathcal{M}_n\in \mathcal{R}^{N_n\times Q}$, where $N_n$ is the number of news and $Q$ is the dimension of the ID embedding. 
The output is $[\mathbf{m}_{n_1}, \mathbf{m}_{n_2}, ..., \mathbf{m}_{n_D}]$.
Upon it, we apply an additive attention layer to combine neighbor ID embeddings into a unified output vector.
The calculation of attention is similar with that in Eq.(\ref{equ:att}).
The final two-hop neighbor news ID representation of news $n$ is denoted as $\mathbf{n}_{e}^{T}$.

\textbf{Neighbor News Semantic Representations.}
Although the ID embeddings of news are unique and inherently represent the neighbor news, they encode news information implicitly.
Moreover, the IDs of some newly-sprung news may not be included in the predefined news ID embedding matrix $\mathcal{M}_n$.
Thus, we propose to attentively learn their context representations via the transformer simultaneously.
For the neighbor news list $[n_{n_1}, n_{n_2}, ..., n_{n_D}]$, the transformer outputs are $[\mathbf{v}_{n_1}, \mathbf{v}_{n_2}, ..., \mathbf{v}_{n_D}]$.
Then the attention layer is applied to model varied importance of neighbor news. 
The final neighbor news semantic representation is the output of the attention layer, which is denoted as $\mathbf{n}_t^T$.

\subsection{Recommendation and Model Training}
The final representations of users and news are the concatenation of outputs from the \textit{one-hop interaction learning} module and the \textit{two-hop graph learning} module, i.e., $\mathbf{u}=[\mathbf{u}_t^{O}; \mathbf{u}_e^{O}; \mathbf{u}_e^{T}]$ and $\mathbf{n}=[\mathbf{n}_t^{T}; \mathbf{n}_e^{T}; \mathbf{n}_t^{O}]$. 
The rating score of a user-item pair is predicted by the inner product of user and item representation, i.e., $\hat{y}=\mathbf{u}^T\mathbf{n}$.
Through this operation, the ID representations and semantic representations are optimized in the same vector space.

Motivated by~\cite{huang2013learning,zhai2016deepintent}, we formulate the click prediction problem as a pseudo $\lambda + 1$ way classification task.
We regard the clicked news as positive and the rest $\lambda$ unclicked news as negative.
We apply maximum likelihood method to minimize the log-likelihood on the positive class:
\begin{equation}
    \mathcal{L}=-\sum_i \log (\frac{\exp (\hat{y}_i^+)}{\exp (\hat{y}_i^+) + \Sigma_{j=1}^\lambda \exp(\hat{y}_{i,j}^-)}),
\end{equation}
where $\hat{y}_i^+$ is the predicted label of the $i_{th}$ positive sample and $\hat{y}_{i,j}^-$ is the predicted label of the associated $j_{th}$ negative sample.
\section{Experiments}
\subsection{Datasets and Experimental Settings}
We constructed a large-scale real-world dataset by randomly sampling user logs from MSN News,~\footnote{https://www.msn.com/en-us/news.} statistics of which are shown in Table~\ref{tab:dataset}.
The logs were collected from Dec. 13rd, 2018 to Jan. 12nd, 2019 and split by time, with logs in the last week for testing, 10\% of the rest for validation and others for training.

In our experiment, we construct $D$ neighbors of the candidate news by random sampling from the clicked logs of its previous users.
For the target user, since there exist massive neighbors users, we rank them according to the number of common clicked news with the target user.
Then we pertain the top $D$ users and use them as graph inputs.
Here we set $D$ to be 15 and use zero padding for cold-start user and newly-sprung news.~\footnote{Due to limit of computational resources,we set $D$ to be this moderate value.}
The dimensions of word embedding, topic embedding and ID embedding are set to 300, 128 and 128 respectively.
We use the pretrained Glove embedding~\cite{pennington2014glove} to initialize the embedding matrix. 
There are 8 heads in the multi-head self-attention network, and the output dimension of each head is 16.
The negative sampling ratio $\lambda$ is set to 4.
The maximum number of user clicked news is set to 50, and the maximum length of news title is set to 30.
To mitigate overfitting, we apply dropout strategy~\cite{srivastava2014dropout} with the rate of 0.2 after outputs from the transformer and ID embedding layers.
Adam~\cite{kingma2014adam} is set to be the optimizer and the batch size is set to be 128. 
These hyperparameters are selected according to the performances on the validation dataset.

For evaluation, we use the average AUC, MRR, nDCG@5 and nDCG@10 scores over all impressions. 
We independently repeat each experiment for 5 times and report the average results.

\begin{table}[t]
    \centering
    \caption{Statistics of our dataset.}
\resizebox{0.48\textwidth}{!}{
    \begin{tabular}[t]{|c|c|c|c|}
         \hline
         \# users&242,175&\# samples&32,563,990\\
         \hline
         \# news&249,038&\# positive samples&805,411\\
         \hline
         \# sessions&377,953&\# negative samples&31,758,579\\
         \hline
         \# avg. words per title&10.99&\# topics&285\\
         \hline
     \end{tabular}
}
    \label{tab:dataset}
\end{table}

\subsection{Performance Evaluation}
In this section, we will evaluate the performance of our approach by comparing it with some baseline methods and a variant of our own method, which are listed as follow:

\begin{itemize}
    \item \textit{NGCF}~\cite{NGCF19}:
    a graph neural network based collaborative filtering method for general recommendation.
    They use ID embeddings as node representations.

    \item \textit{LibFM}~\cite{rendle2012factorization}:
    a feature based model for general recommendation using matrix factorization.
    % TF-IDF~\cite{jones2004statistical} features are used as representations.
    % We extract the TF-IDF~\cite{jones2004statistical} feature from the concatenation of the clicked or candidate news titles and categories as sparse user features.

    % \item \textit{DeepFM}~\cite{guodeepfm}: 
    % a general deep learning based factorization model. 
    % We also use the TF-IDF text feature as the input.

    \item \textit{Wide\&Deep}~\cite{cheng2016wide}: 
    a general recommendation model which has both a linear wide channel and a deep dense-layer channel. 
    % We use the TF-IDF text feature mentioned above as the input.

    \item \textit{DFM}~\cite{lian2018towards}: 
    a neural news model utilizing an inception module to learn user features and a dense layer to merge them with item features.
    % TF-IDF

    \item \textit{DSSM}~\cite{huang2013learning}: 
    a sparse textual feature based model which learns news representation via multiple dense layers.
    
    \item \textit{DAN}~\cite{zhu2019dan}: 
    a CNN based news model which learns news representations from news titles.
    An attentional LSTM is used to learn user representations.
    
    \item \textit{GRU}~\cite{okura2017embedding}: 
    a deep news model using an auto-encoder to learn news representations and a GRU network to learn user representations.
    
    \item \textit{DKN}~\cite{wang2018dkn}: 
    a CNN based news model enhanced by the knowledge graph.
    They utilize news-level attention to form user representations.
    
    \item \textit{GERL-Graph}:
    Our model without the \textit{two-hop graph learning}.
\end{itemize}

\begin{table}[t]
  \caption{The performance scores and standard variations of different methods. *The improvement is significant at the level p < 0.002.}
  \label{tab:baseline}
  \resizebox{0.48\textwidth}{!}{
  \begin{tabular}{ccccc}
    \toprule
    Methods&AUC&MRR&nDCG@5&nDCG@10\\
    \midrule
    NGCF~\cite{NGCF19}&55.45$\pm$0.16&17.19$\pm$0.05&17.23$\pm$0.10&22.08$\pm$0.09\\
    LibFM~\cite{rendle2012factorization}&61.83$\pm$0.10&19.31$\pm$0.06&20.45$\pm$0.08&25.69$\pm$0.08\\
    Wide\&Deep~\cite{cheng2016wide}&64.62$\pm$0.14&20.71$\pm$0.12&22.43$\pm$0.15&27.99$\pm$0.15\\
    DFM~\cite{lian2018towards}&64.72$\pm$0.19&20.75$\pm$0.14&22.60$\pm$0.20&28.22$\pm$0.19\\
    DSSM~\cite{huang2013learning}&65.49$\pm$0.18&20.93$\pm$0.13&22.93$\pm$0.22&28.65$\pm$0.27\\
    DAN~\cite{zhu2019dan}&65.52$\pm$0.13&21.25$\pm$0.18&23.14$\pm$0.21&28.73$\pm$0.15\\
    GRU~\cite{okura2017embedding}&65.69$\pm$0.19&21.29$\pm$0.10&23.16$\pm$0.11&28.75$\pm$0.11\\
    DKN~\cite{wang2018dkn}&65.88$\pm$0.13&21.46$\pm$0.21&23.23$\pm$0.25&28.84$\pm$0.21\\
    \midrule
    GERL-Graph&67.74$\pm$0.13&22.71$\pm$0.15&25.03$\pm$0.13&30.65$\pm$0.15\\
    \midrule
    \textbf{GERL}&\textbf{68.55$\pm$0.12}&\textbf{23.33$\pm$0.10}&\textbf{25.82$\pm$0.14}&\textbf{31.44$\pm$0.12}\\
  \bottomrule
\end{tabular}
 }
\end{table}

For fair comparison, we extract the TF-IDF~\cite{jones2004statistical} feature from the concatenation of the clicked or candidate news titles and topics as sparse feature inputs for LibFM, Wide\&Deep, DFM and DSSM.
For DSSM, the negative sampling ratio is also set to 4.
% we used the concatenation of the clicked news titles as the query and the titles of the candidate news as the document. 
We try to tune all baselines to their best performances.
The experimental results are summarized in Table~\ref{tab:baseline}, and we have several observations:

First, methods which represent news directly from news texts (e.g., DAN, GRU, DKN, GERL-Graph, GERL) usually outperform feature based methods (e.g., LibFM, Wide\&Deep, DFM, DSSM).
The possible reason is that although feature based methods learn news content, the useful information exploited from news texts is limited, which may lead to sub-optimal news recommendation results.

Second, compared with NGCF, which also exploits neighbor information in the graph, our method achieves better results.
This is because NGCF is an ID-based collaborative filtering method, which may suffer from cold-start problem significantly.
% ~\footnote{There are only 6375 (2.63\%) overlap users between our training and testing set.}
This result further proves the effectiveness of introducing textual understanding into graph neural networks for news recommendation.

Third, compared with other methods that involve textual content of news (e.g., DAN, GRU, DKN), our GERL-Graph can consistently outperform other baseline methods.
This may because the multi-head attention in transformer module learns contextual dependency accurately.
Moreover, our approach utilizes attention mechanism to select important words and news.

Fourth, our GERL approach which combines both textual understanding and graph relatedness learning outperforms all other methods.
This is because GERL encodes neighbor user and news information by attentively exploiting the interaction graph.
The result validates the effectiveness of our approach.

\subsection{Effectiveness of Graph Learning}
\begin{figure}[t]
    \centering
    \resizebox{0.37\textwidth}{!}{\includegraphics{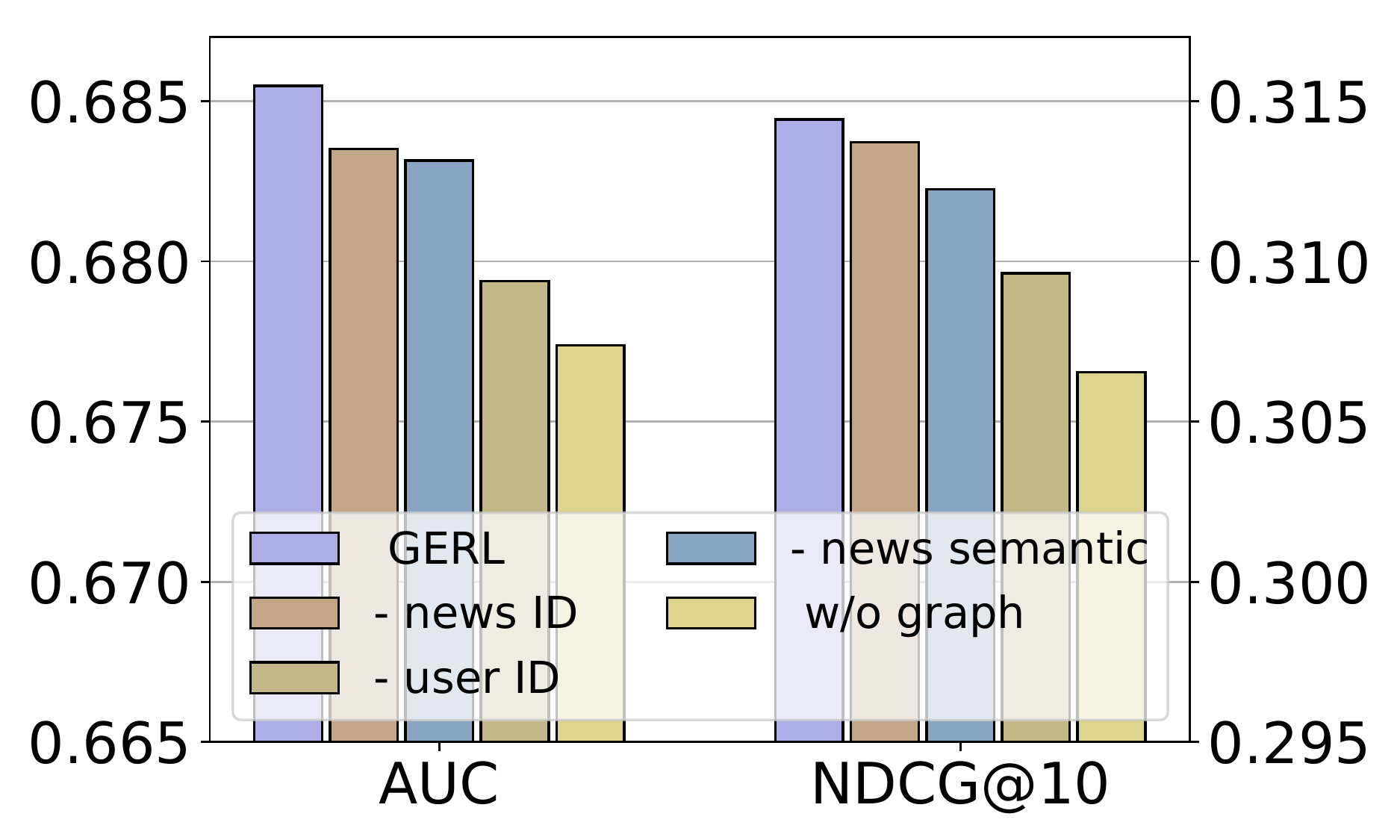}}
    \caption{Effectiveness of two-hop graph learning.}
    \label{fig:grapheffect}
\end{figure}

To validate the effectiveness of the \textit{two-hop graph learning} module, we remove each component of representations in the module to examine its relative importance and illustrate the results in Figure~\ref{fig:grapheffect}.~\footnote{We use a trainable dense layer to transform vector $\mathbf{u}$ or $\mathbf{v}$ and keep the dimension uniform as before.}
% From the figure, we can conclude that each type of second-order relationship help our model learn better user and news representations for recommendation.
% Among three kinds of graph interaction, 
% \textit{User ID > News content > News ID}
% in effectiveness level.
% Several reasons may explain for this.
% Several observations can be made from it.
Based on it, several observations can be made.
First, adding the neighbor user information improves performances more significantly than adding neighbor news information.
In our GERL-Graph approach, candidate news can be directly modeled through titles and topics, while target users are only represented by their clicked news.
When the user history is sparse, they may not be well represented.
Hence, adding IDs of neighbor users may assist our model to learn better user representations.
Second, the improvement brought by neighbor news semantic representations outweighs that brought by neighbour news ID.
This is intuitive since titles of news contain more explicit and concrete meanings than IDs.
Third, combining each part in the graph learning leads to the best model performance.
By adding graph information both from neighbor users and news, our model forms better representations for recommendation.

\subsection{Ablation Study on Attention Mechanism}
\begin{figure}[t]
    \centering
    \quad
    \subfigure[Transformer attention.]{\label{fig:transatt} \resizebox{0.22\textwidth}{!}{\includegraphics{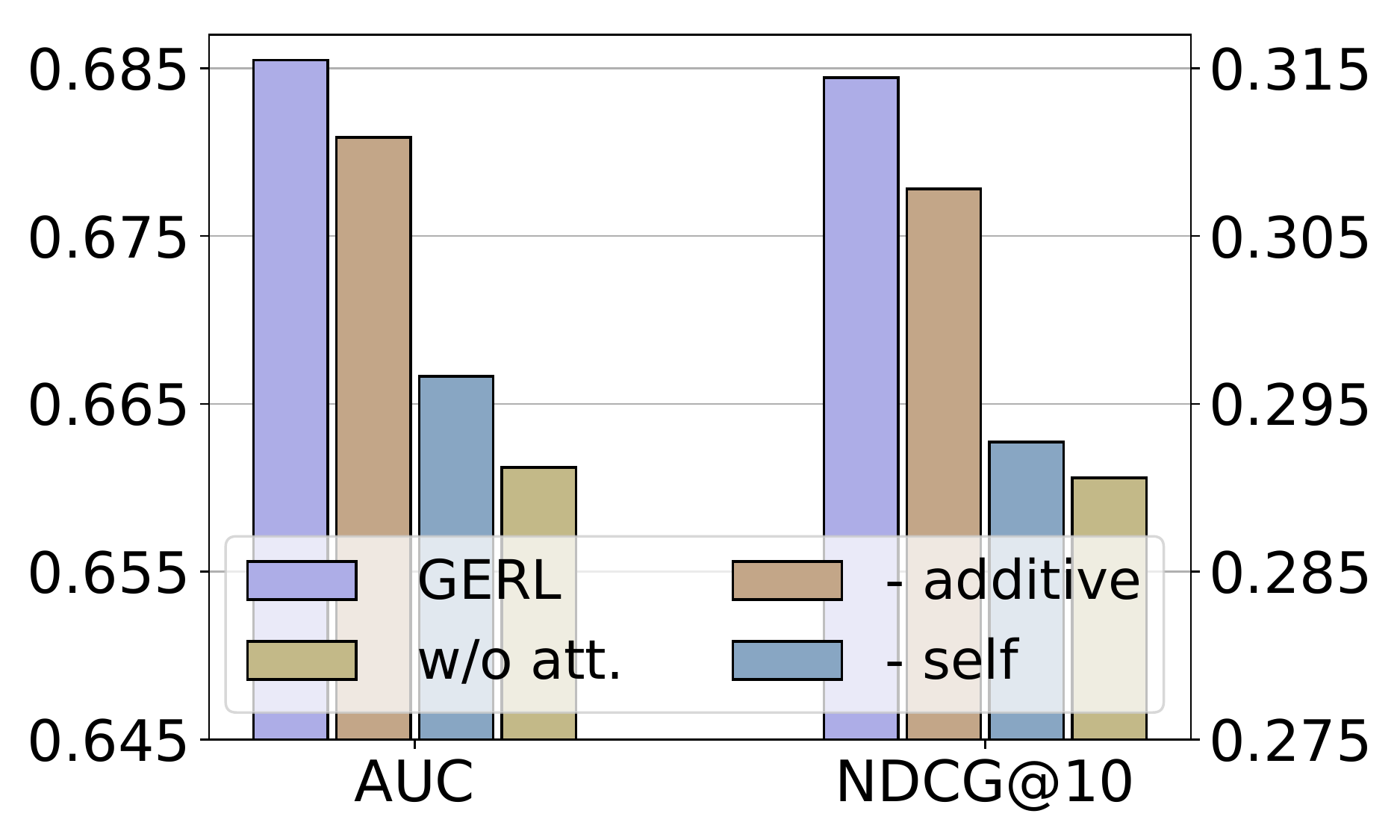}} }
    \subfigure[Model attention.]{\label{fig:modelatt} \resizebox{0.22\textwidth}{!}{\includegraphics{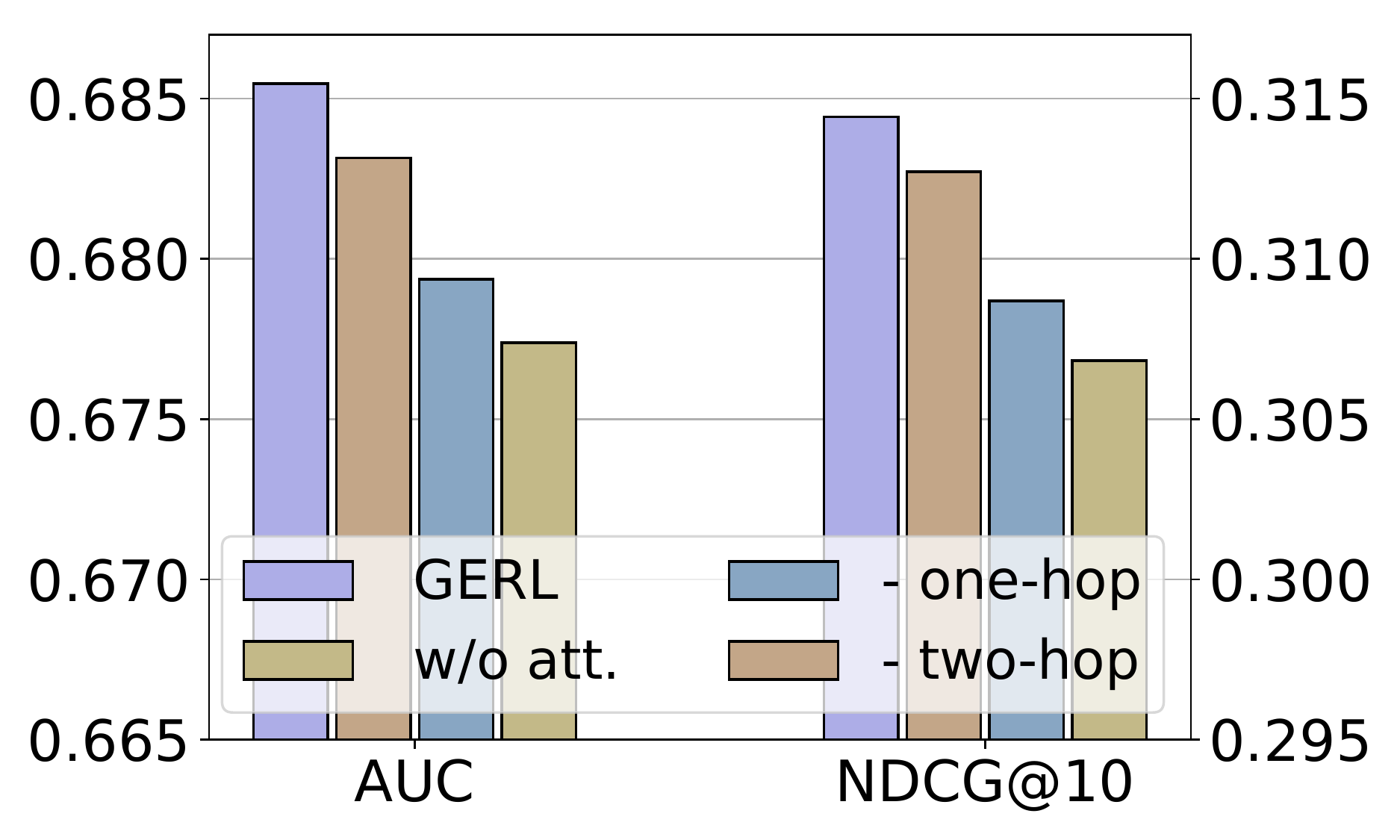}}}
    \caption{Effectiveness of attention mechanism.}
    \label{fig:att}
\end{figure}

Next, we explore the effectiveness of two categories of attention by removing certain part of them. 
Instead, to keep dimensions of vectors unchanged, we use average pooling to aggregate information.
First, we verify two types of attention inside the transformer in Figure~\ref{fig:transatt}.
From it, we conclude that both the additive and the self attention are beneficial for news context understanding.
This is because self-attention encodes interactions between words and additive attention helps to select important words.
Among them, self-attention contributes more to improving model performances, as it models both short-distance and long-distance word dependency.
Moreover, it forms diverse word representations with multiple attention heads.
Also, we verify the model-level attention, e.g., attention inside the \textit{one-hop interaction learning} and that in the \textit{two-hop graph learning}.
From Figure~\ref{fig:modelatt}, we observe that the  attention in the one-hop module is more important.
One-hop attention selects important clicked news of users, thus helping model user preferences directly.
Compared with that, two-hop attention models relative importance of neighbors, which may only represent interests implicitly.
By using both attentions simultaneously, we obtain the best performances.

\subsection{Hyperparameter Analysis}

Here we explore the influences of two hyperparameters.
One is the number of attention heads in the transformer module.
Another one is the degree of graph nodes in the graph learning module.

\begin{figure}[t]
    \centering
    \subfigure[Attention head number.]{\label{fig:atthead} \resizebox{0.197\textwidth}{!}{\includegraphics{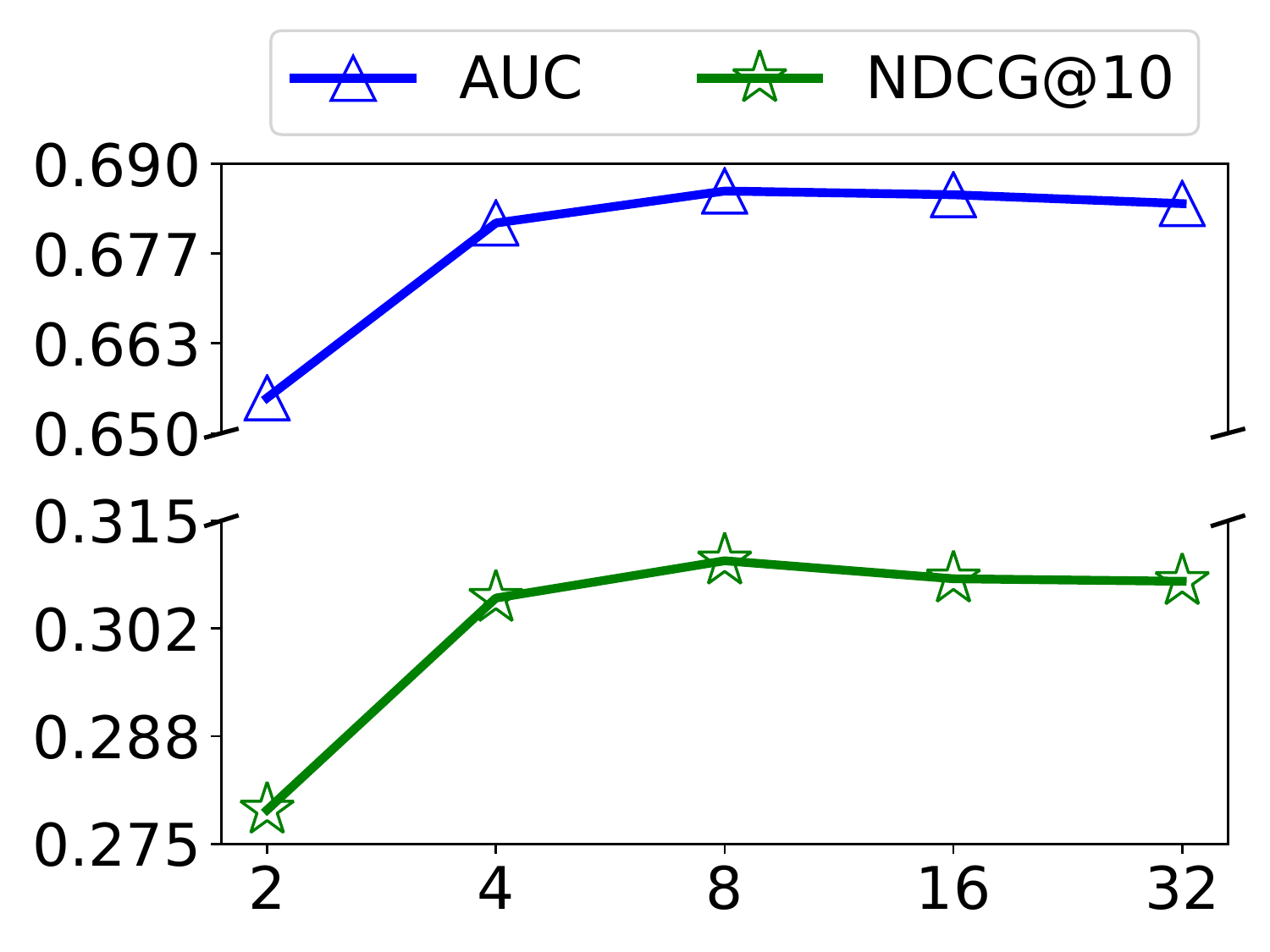}} }
    \subfigure[Node degree.]{\label{fig:degree} \resizebox{0.2\textwidth}{!}{\includegraphics{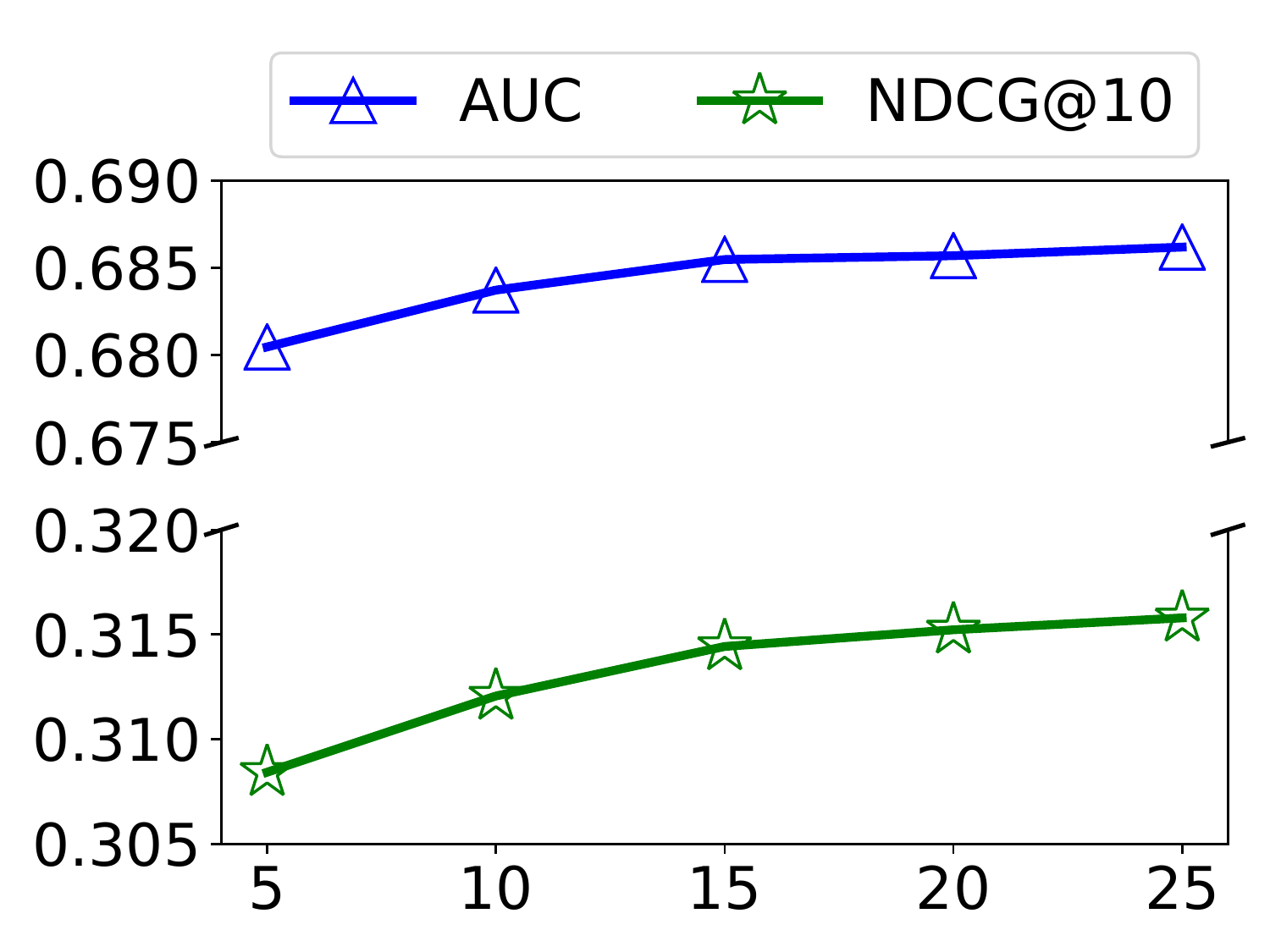}}}
    \caption{Influence of two hyperparameters.}
    \label{fig:hyper}
\end{figure}

\textbf{Number of Attention Heads.}
In the transformer module, the number of self-attention heads is crucial for learning context representations.
We illustrate its influence in Figure~\ref{fig:atthead}.
An evident increase can be observed when the number increases from $2$ to $8$, as the rich textual meanings may not be fully exploited when there are few heads.
However, the performances drop a little when head number increases from 8.
This may happen because news titles are concise and brief, thus too many parameters may be sub-optimal.
Based on the above discussion, we set the number to be 8.

\textbf{Degree of graph nodes.}
In the graph learning module, the degree of user and item nodes decides how many similar neighbors our model will learn.
We increase the node degree from $5$ to $25$ and showcase its influence in Figure~\ref{fig:degree}.
As illustrated, the performance improves when more neighbors are taken as model inputs, which is intuitive because more relatedness information from the graph is incorporated.
Meanwhile, the increasing trend becomes smooth when the degree is larger than $15$.
Therefore, we choose a moderate value $15$ as the number of node degree.

\section{Conclusion}
In this paper, we propose a graph enhanced representation learning architecture for news recommendation.
% Our approach combines context understanding and graph neighbor relatedness learning in a unified manner.
% Under this framework, semantic news representations are formed from their titles and topics via a transformer architecture.
% To further introduce neighbor news information, instead of recursively propagating on the whole graph, we aggregate the semantic and ID representations of neighbors via a graph attention network.
% When learning user representations, we incorporate their clicked news representations via attention.
% Besides, ID embeddings from neighbor users in the graph is utilized simultaneously.
Our approach consists of a \textit{one-hop interaction learning} module and a \textit{two-hop graph learning} module.
The \textit{one-hop interaction learning} module forms news representations via the transformer architecture.
It also learns user representations by attentively aggregating their clicked news.
The \textit{two-hop graph learning} module enhances the representations of users and news by aggregating their neighbor embeddings via a graph attention network.
Both IDs and textual contents of news are utilized to enrich the neighbor embeddings.
% Performance improvement on a real-world dataset proves the effectiveness of our approach.
Experiments are conducted on a real-world dataset, the improvement of recommendation performances validates the effectiveness of our approach.

\begin{acks}
  The authors would like to thank Microsoft News for providing technical support and data in the experiments. This work was supported by the National Key Research and Development Program of China under Grant number 2018YFC1604002, the National Natural Science Foundation of China under Grant numbers U1836204, U1705261, U1636113, U1536201, and U1536207, and the Tsinghua University Initiative Scientific Research Program under Grant number 20181080368.
\end{acks}

\bibliographystyle{ACM-Reference-Format}
\balance
\bibliography{main}
\end{document}